\documentclass[epjST]{svjour}
\usepackage{graphics}
\newcommand{\beq}{\begin{equation}}
\newcommand{\eeq}{\end{equation}}
\newcommand{\beqa}{\begin{eqnarray}}
\newcommand{\eeqa}{\end{eqnarray}}
\newcommand{\ba}{\begin{array}}
\newcommand{\ea}{\end{array}}
\usepackage[dvips]{epsfig}

\begin{document}

\title{Two-mode dipolar bosonic junctions}

\author{G. Mazzarella and L. Dell'Anna}

\institute{Dipartimento di Fisica e Astronomia ``Galileo Galilei''
and CNISM, Consorzio Nazionale Interuniversitario per le Scienze Fisiche della Materia (CNISM), Universit\`a di Padova, Via Marzolo 8, I-35131 Padova, Italy}

\date{\today}

\abstract{We consider a two-mode atomic Josephson junction realized with dilute dipolar bosons confined by a double-well.
We employ the two-site extended Bose-Hubbard Hamiltonian and characterize
the ground-state of this system by the Fisher information, coherence visibility, and entanglement entropy. These quantities are
studied as functions of the interaction between bosons in different wells.
The emergence of Schr\"odinger-cat like state with a loss of coherence is
also commented.}


\maketitle

\section{Introduction}

Ultracold and dilute bosons confined in one-dimensional (1D) double-well potentials \cite{oliver}
are the ideal framework to address the study the Josephson effect \cite{book-barone}, and the formation of macroscopic coherent states \cite{smerzi,stringa,anglin,mahmud,anna} and Schr\"odinger-cat states \cite{cirac,dalvit,huang,carr,brand,cats,dellanna}. The coherent dynamics of the bosonic cloud in the double-well potential (bosonic Josephson junction) \cite{oliver,exp-bec}, widely studied with alkali-metal atoms, is very well described by Josephson equations \cite{smerzi}, and their extensions \cite{bergeman,ajj1,ajj2,sb}. The Josephson equations are valid when the inter-atomic interaction is weak against the Josephson coupling energy (i.e., the tunneling probability amplitude multiplied by the number of bosons) and the number of bosons is much larger than one. In this case, semiclassical approximations can
be performed and the system is considered in a coherent state \cite{leggett}.
By increasing the coupling strength of the boson-boson repulsive interaction
\cite{stringa,anglin,mahmud,anna} the crossover from a coherent
state (superfluid-like regime) to a pure Fock state (Mott-like regime) takes place.
For attractive bosons, the Josephson equations predict the
spontaneous symmetry-breaking when the strength is above a critical value \cite{smerzi,sb}, while the two-site Bose-Hubbard model \cite{milburn} predicts the formation of a Schr\"odinger-cat state \cite{cirac,dalvit,huang,carr,brand,cats,dellanna}. When the attraction between
the bosons becomes strong enough, one expects the collapse of the cloud \cite{sb,luca-e-boris}.

The above considerations hold for bosons with negligible dipole or electric moments, so that the dipole-dipole interatomic interaction can be be safely neglected. This is not the case for junctions made by dipolar bosons, as for example $^{52}$Cr, characterized by very large atomic magnetic dipoles.

In the present contribution we study the emergence of cat-like states in atomic Josephson junctions made of dipolar bosons. Recently, dipolar quantum gases, where considering long-range and anisotropic dipole-dipole interaction  between magnetic or electric dipoles makes sense, have attracted a lot of interests \cite{baranov,lahaye}. To date, important theoretical efforts have been devoted in order to investigate dipolar bosonic gases trapped in double \cite{xiong,blume,abad} and triple wells \cite{lahaye2}. Remarkably, Abad and co-workers \cite{abad} have shown that the dipolar interaction makes possible self-inducing a double-well potential structure.

We use as theoretical tool the extended two-site Bose-Hubbard (EBH) Hamiltonian \cite{lahaye2,ebh,goral} where both the on-site (intra-well) interaction and nearest-neighbor (inter-well) density-density one are considered.
We follow the same path as in \cite{bruno,cats,dellanna}. Thus, we diagonalize the two-site EBH Hamiltonian and study the Fisher information $F$, the coherence visibility $\alpha$, and the entanglement entropy $S$ of the ground-state
by fixing the on-site interaction amplitude and varying the inter-well density-density interaction.
We find that the presence of a  macroscopic superposition state corresponds to a sufficiently large values of the Fisher information $F$, as expected from the studies presented in \cite{lorenzo,pezze}. By increasing the inter-well interaction, $F$ grows, the coherence visibility goes to zero, and the entanglement entropy $S$ reaches a maximum value - in correspondence to a given inter-well interaction.
For larger nearest-neighbor interaction, $S$ decreases and the ground-state of the EBH evolves toward a cat-like state.

As a second task, we solved the ordinary differential equations
for a bosonic junction obtained by using the quasi-classical coherent state,
varying the inter-well interaction. When such an interaction becomes strong enough, the population imbalance starts to oscillate around a non-zero value. The on-set of this behavior takes place for that inter-well interaction corresponding to the maximum value of the entanglement entropy $S$.

\section{The model Hamiltonian}

We consider an ultracold gas of $N$ identical dipolar bosons of mass $m$. The boson-boson interaction derives from the sum of a short-range contact potential
and a long-range dipole-dipole potential, i.e. $\displaystyle{V({\bf r}-{\bf r}')=g\delta({\bf r}-{\bf r}')+\gamma \frac{1-3 \cos^2 \theta}{|{\bf r}-{\bf r}'|^3}}$,
where $g=4\pi\hbar^2 a_s/m$ with $a_s$ the interatomic s-wave scattering length; $\gamma=\mu_0\mu^2/4\pi$ for magnetic dipoles ($\mu_0$ is the vacuum magnetic susceptibility and $\mu$ is the magnetic dipole moment) or $\gamma=d^2/4\pi \varepsilon_0$ for electric dipoles ($\varepsilon_0$ is the vacuum dielectric constant and $d$ is the electric dipole moment). We assume that the bosons are polarized by a sufficiently large external field with all the dipoles aligned along the same direction; $\theta$ is the angle between the vector ${\bf r}-{\bf r}'$ and the dipole orientation.
We suppose the bosons are confined by the superposition of an isotropic harmonic confinement in the transverse ($y-z$) radial plane and a symmetric double-well potential $V_{DW}$ in the axial direction ($x$):
$V_{trap}({\bf r})= V_{DW}(x)+\frac{1}{2}m\omega_{\bot}^2(y^2+z^2)$
with $\omega_{\bot}$ the trapping frequency in the radial plane. In the following we shall consider a strong transverse confinement so that the system can be treated as one-dimensional (1D). In particular, the transverse energy $\hbar \omega_{\bot}$ is assumed much larger than the characteristic energy of bosons in the axial direction.
Then, a dilute bosonic gas 
can be described with the help of the two-mode Bose-Hubbard Hamiltonian \cite{milburn}. We are considering particles that interact with each other via long-range forces; thus the interactions between bosons in different wells have to be taken into account. Proceeding from the second quantized Hamiltonian written in terms of the space dependent bosonic field operators, one integrates out the spatial degrees of freedom, namely ${\bf r}$ and ${\bf r'}$, involved in the trapping and boson-boson interaction potentials and in the kinetic term as well. Thus one obtains the following two-site version of extended Bose-Hubbard (EBH) Hamiltonian \cite{goral} (in the following $L$ will stand for left, $R$ for right):
\beq
\label{twomode}
\hat{H} = -J\big(\hat{a}^{\dagger}_L\hat{a}_R
+\hat{a}^{\dagger}_R\hat{a}_L\big)
+\frac{U_0}{2}\big( \hat{n}_L (\hat{n}_L -1)
+ \hat{n}_R (\hat{n}_R -1)\big)+
U_1 \hat{n}_L \hat{n}_R \;,
\eeq
where $\hat{a}_{k},\hat{a}^{\dagger}_{k}$ ($k=L,R$) are bosonic operators
and $\hat{n}_{k}=\hat{a}^{\dagger}_{k}\hat{a}_{k}$ counts the number of particles in the $k$th well. Notice that due to the above mentioned integration over the spatial degrees of freedom, the quantities $J$, $U_0$, and $U_1$ do not depend on the spatial coordinates, but only on the microscopic parameters of the system, see, for example
\cite{ajj2}. In particular, $J$ is the hopping amplitude between the two wells,
$U_0$ the on-site interaction amplitude, and $U_1$ the nearest-neighbor interaction amplitude.

\section{Analysis}

In the absence of the nearest-neighbor interaction ($U_1=0$), the ground-state of the Hamiltonian (\ref{twomode}) in the limit $U_0/J \rightarrow -\infty$ is the macroscopic superposition state
$\frac{1}{\sqrt{2}}(|N,0\rangle+|0,N\rangle)$,
also known as "NOON" state or "Schr\"odinger cat state". However, when the interaction between the bosons is attractive, i.e. $a_s<0$ (resulting in $U_0/J<0$), the collapse of the atomic cloud can be observed if the atomic density is sufficiently high. Apart the question of the possible collapse, the realization of the cat state is not trivial due to the very tiny separation (in the presence of finite couplings)  between the two lowest levels that makes the cat state very fragile, see, for example \cite{huang}.


Let us suppose that $U_1 \neq 0$ and $J=0$. It is easy to see that, for $U_1>U_0$, the Hamiltonian (\ref{twomode}) admits the NOON state as its lowest eigenstate even if the on-site interaction is repulsive. This allows to realize cat-like states by-passing the collapse problem. It is worth to observe that, in the absence of the hopping, the emergence of the NOON state can be understood in terms of the interplay between the intra-well interaction, in practice $U_0 \hat{n}_{k}^{2}$ ($k=L,R$), and the inter-well one, i.e.  $U_1 \hat{n}_L\hat{n}_R$. In fact, the effect of the former is to establish a balanced population among the two sites, while the latter is minimum when one of the on-site average occupations vanishes.

We therefore study the emergence of cat-like state by fixing $U_0$ and varying $U_1$ for a given $N$. This is possible since - see, f.i., \cite{lahaye2} - the on-site interaction results from short-range interaction and dipole-dipole one, while $U_1$ depends only on the dipole-dipole interaction: then, we have to vary both contact and dipole-dipole interaction simultaneously, in such a way that we can change $U_1$ keeping $U_0$ constant.

Since the total number of bosons is a preserved quantity, we numerically solve the eigenproblem
${\hat H} |E_j\rangle = E_j |E_j \rangle$
for a fixed number $N$ of bosons.
For each eigenvalue $E_j$ ($E_0<E_1<E_2<...$) with $j=0,1,...,N$, the corresponding eigenstate $|E_j\rangle$ will be of the form
\beq
|E_j\rangle=\sum_{i=0}^{N}\,c_{i}^{(j)} \, |i,N-i\rangle \;
\label{eigenstate}
\;.\eeq
\begin{figure}[ht]
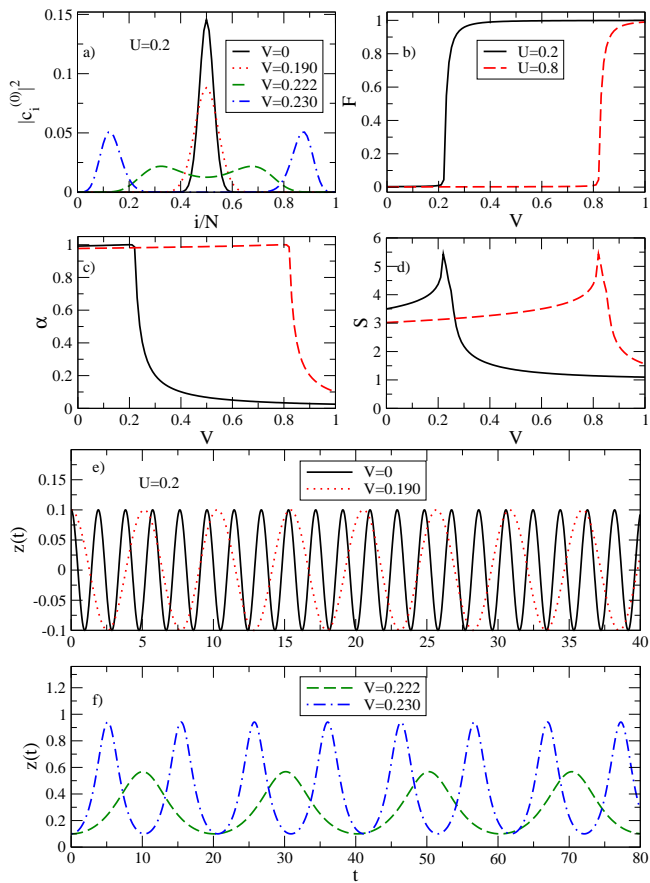

\begin{center}
\epsfig{file=fig2gd.eps,width=0.65\linewidth,clip=}
\epsfig{file=fig1lion.eps,width=0.65\linewidth,clip=}
\end{center}
\caption{(Color online). a) Coefficients $|c_{i}^{(0)}|^2$ for
the ground-state as a function of $i/N$ for $U=0.2$. b) The normalized quantum Fisher information $F$ vs $V$ at $U=0.2$ (solid line) and $U=0.8$ (dotted line). c) The coherence visibility $\alpha$ vs $V$ at $U=0.2$ (solid line) and $U=0.8$ (dotted line).
d) The entanglement entropy $S$ vs $V$ at $U=0.2$ (solid line) and $U=0.8$ (dotted line).
e) and f) The population fractional imbalance $z(t)$ vs time $t$ at $U=0.2$. Initial conditions $z(0)=0.1$, $\phi(0)=0$. $|c_{i}^{(0)}|^2$,
$i/N$, $U$, and $V$ are all dimensionless quantities. Number of bosons $N=100$. Time is in units of $\hbar/J$.}
\label{fig1}
\end{figure}
Let us fix the rescaled on-site interaction $U=U_0/J$ at a positive value and vary the rescaled nearest-neighbor interaction $V=U_1/J$. We find that it exists a crossover value of $V$, say $V_{cr}$. When $V$ exceeds such a value, the system begins to lose the coherence. This crossover connects a ground-state with the maximal probability at $i=N/2$ (single peak centered around $i=0$) - for $V=0$ - to a ground-state which has the maximal probability at $i=n$ and $i=N-n$ (two separated peaks symmetric with respect to $i=0$) with $n$ approaching $0$ as $V$ increases (this last situation corresponds to the emergence of the Schr\"odinger cat state), as it can be clearly observed from the panel a) of Fig. 1, where we have plotted the coefficients $|c_{i}^{(0)}|^2$. We have numerically found that the $V$ crossover value is related to $U$ by the following linear relation:
\beq
\label{linearrelation}
V_{cr}=U+A
\;,\eeq
where $A>0$ depends on $N$. For example, when $N=30$, $A=0.075$, while when $N=100$, $A$ is equal to $0.0212$.
Translated into the language of the atomic Josephson junctions: when $V$ approaches $V_{cr}$ the junction is entering the self-trapping regime
in which the fractional population imbalance between the two wells, $z=(N_L-N_R)/N$ ($N_k$ is the number of bosons in the $k$th well)
oscillates around a non-zero time averaged value, see the panel f) of Fig. 1
and the discussion below.

From an analytical point of view, the above quoted interpretation can be supported by calculating the expectation value $E$ of the EBH Hamiltonian (\ref{twomode}) with respect to the state \cite{cats}
$|QC\rangle=|CS\rangle_{L} \otimes |CS\rangle_{R}$,
where $|CS\rangle_{k}$ ($k=L,R$) - which describes the Bose-Einstein condensate (BEC) in the $k$th well - is \cite{glauber}
\beq
\label{CSI}
|CS\rangle_{k}=e^{-|z_k|^{2}/2}\sum_{n=0}^{\infty}\frac{z_{k}^n}
{\sqrt{n!}}|n\rangle \;.
\eeq
The complex quantity $z_k$ is the eigenvalue of the annihilator in the
$k$th well, i.e.
\beq
\label{zetai}
\hat{a}_k |CS\rangle_{k}=z_k\,|CS\rangle_{k} \; .
\eeq
The absolute values of the two $z_k$ are related to the average
occupancy of the two wells:
\beq
N_k\equiv \langle QC|{\hat n}_k|QC\rangle = |z_k|^2  \; .
\eeq
Therefore $z_k$ are conveniently parametrized as
$z_k=\sqrt{N_k}\exp(i\theta_k)$, where $\theta_k$ are
phase variables.
Following the same procedure as in \cite{cats} when only $U_0$ was present, we get for the expectation value $E$ the following expression:
\beq
\label{e}
E=\langle QC|{\hat H}|QC\rangle =
-N\,J\sqrt{1-z^2}\cos \phi
+\frac{N^2\,U_{0}}{4}\,(1+z^2)+\frac{N^2\,U_{1}}{4}\,(1-z^2)
\;,\eeq
where $\phi=\theta_R-\theta_L$ ($\theta_k$ is the phase of the BEC in the $k$th well) and the fractional imbalance $z=(|z_L|^2-|z_R|^2)/(|z_L|^2+|z_R|^2)$. Starting from the energy (\ref{e}), the equations of the motion for the generalized coordinate $\phi$ and its conjugate moment $z$ 
provide the ordinary differential equations (ODEs) for $z(t)$ and $\phi(t)$. These ODEs, with the time scaled with respect to $\hbar/J$, read
\beqa
\label{odes}
&&\dot{z}(t)=
-\sqrt{1-z^2(t)}\,\sin \phi(t)\nonumber\\
&&\dot{\phi}(t)=
\frac{U-V}{2}N z(t)+\frac{z(t)\cos\phi(t)}{\sqrt{1-z^2(t)}}.\;\eeqa
By solving these ODEs, we have obtained the panels e) and f) of Fig. 1. Now, it is straightforward to show that the non-zero minimum of the energy (\ref{e}) is obtained with $\phi=0$ and
\beq
\label{minima}
z=z_{sb}=\pm \sqrt{1-\frac{4}{(U-V)^2\,N^2}}\,,\;\;
\quad \mbox{if} \quad  (V-U)> \frac{2}{N}
\;.\eeq
It is easy to show that stationary points of the form $(z,0)$ (with $z \neq 0$ and zero relative phase $\phi$) for the ordinary differential equations (\ref{odes}) can exist only when $V>U$ ($U_1>U_0$). In particular, the point $(z_{sb},0)$, see Eq.(\ref{minima}), is a stationary point for the afore mentioned ODEs if the order relation in Eq.(\ref{minima}) is met. Notice that if the ODEs (\ref{odes}) are solved with the initial condition $\phi(t=0) \equiv \phi(0)=0$, oscillations of the population imbalance around a non-zero time averaged value ($\langle z(t) \rangle \neq 0$) are possible for any $z(t=0) \equiv z(0)$ provided that $V>U+2/N$ ($U_1>U_0+2J/N$). For this kind of oscillations, the temporal average of the relative phase is zero.
On the other hand, if $U>V$ - i.e. $U_0>U_1$ - (and hence, in particular, in the ordinary BJJ case corresponding to $U_1=0$), for $\phi(0)=0$, a non-zero $z$ cannot be a stationary point for the above ODEs. In this case, the oscillations around $\langle z(t) \rangle \neq 0$ take place if $z(0)$ exceeds a critical value and the phase is an increasing function of the time, and we have the so called running phase mode \cite{smerzi}.

At this point it is worth to comment about the following. The $z$-symmetry broken states that one observes around stationary stables point of (\ref{odes}) are related to the semiclassical limit that we have used above. Such states have a large lifetime that scales exponentially with $N$ \cite{smerzi}. In a full quantum two-mode approximation, instead, the ground state is always symmetric in the population imbalance, being well approximated by a symmetric superposition of two unbalanced coherent like states, namely a Schr\"odinger cat state \cite{dellanna}.

We see that at varying $U$ (that is, the intra-well interaction), $V=U+2/N$ - that is nothing but Eq. (\ref{linearrelation}) - gives the value of $V$ (that is, the inter-well interaction) which signals the onset of the self-trapping regime in the bosonic Josephson junction.

By following \cite{cats}, we characterize the ground-state of our system by studying three indicators: the quantum Fisher information, the coherence visibility, and the entanglement entropy. We shall analyze these quantities as functions of $V$ in correspondence to the same on-site rescaled interaction.

It is useful to recall the definition of the quantum Fisher information (QFI) that we shall use in the following: $F_{QFI}=\langle (\hat{n}_{L}-\hat{n}_{R})^{2}\rangle -\big(\langle\hat{n}_{L}-\hat{n}_{R}\rangle\big)^2$ \cite{braunstein}.
The quantum Fisher information is one of the key quantities of the estimation theory \cite{helstrom}. In particular, the QFI is related to the bound on the precision with which the accumulated phase in an interferometric experiment can be determined. For a given input state of an interferometric procedure, the best achievable precision is the reciprocal of the squared root of QFI, known as the quantum Cram\'er-Rao bound \cite{helstrom}. Notice that the definition of QFI above provided is relevant for an atomic interferometer based on rotations of the input state about the $z$ axes of the Bloch sphere. However, other choices would be possible, namely those relying on input state rotations about the $x$ and $y$ axes of the Bloch sphere \cite{pezze,braunstein}. Moreover, the quantum Fisher information is a parameter related to the multiparticle entanglement, namely to the indistinguishability of the bosons \cite{pezze}.


It is convenient to normalize $F_{QFI}$ at its maximum value $N^2$ - as well known from \cite {lorenzo,pezze} - by defining
the normalized quantum Fisher information $F={F_{QFI}}/{N^2}$.
In terms of the coefficients $c_{i}^{(0)}$, $F$ is given by:
\beq
\label{qfigs}
F =\frac{1}{N^2}
\sum_{i=0}^{N}\big[2i-N\big]^{2}|c_{i}^{(0)}|^2 \; .
\eeq
In the large $N$ limit we can compare our results to analytical asymptotic
behaviors \cite{dellanna}. For $V> U+2/N$, in fact, we expect to have
\beq
F(V)\simeq \frac{1}{N}\left(1+(N-1)z_{sb}^2\right)\,,
\eeq
with $z_{sb}$ given by Eq.~(\ref{minima}). In the extreme NOON state ($z_{sb}=1$) we obtain $F=1$. For $U>V$, instead, we get the following large $N$ limit result
\beq
F(V)\simeq \frac{1}{N}\sqrt{\frac{2}{2+(U-V)N}}\;.
\eeq
Let us move, now, to the coherence visibility $\alpha$ given by
$\alpha={2\,|\langle \hat{a}^{\dagger}_L\hat{a}_R\rangle|}/{N}$ \cite{stringa}
which characterizes the degree of coherence between the two wells.
The expectation value of the operator $\hat{a}^{\dagger}_L\hat{a}_R$
is evaluated in the ground-state $|E_0\rangle$ and the visibility $\alpha$ is given by
\beq
\label{jpjm}
\alpha ={2\over N} \sum_{i=0}^{N}
c_{i}^{(0)} c_{i+1}^{(0)}\sqrt{(i+1)(N-i)} \;.
\eeq
In the large $N$ limit and for $V>U+2/N$, the asymptotic behavior is given by
\beq
\alpha(V)\simeq \sqrt{1-z_{sb}^2}\;,
\eeq
while its value for $U>V$, always for large number of bosons, is well approximated by \cite{dellanna}
\beq
\alpha(V)\simeq 1-\frac{1}{2}\sqrt{\frac{(U-V)}{2N}}\,.
\eeq
Finally, we calculate the entanglement entropy $S$ \cite{bwae}. This is an excellent measure of the quantum entanglement of the ground-state $|E_0\rangle$. When the system is in $|E_0\rangle$, the density matrix is
$\hat{\rho} =|E_0\rangle\langle E_0|$.
$S$ is defined as the von Neumann entropy of the reduced density matrix $\hat{\rho}_{L}=Tr_{R} \hat{\rho}$,
which is the matrix obtained by partial tracing the total density matrix $\hat{\rho}$ over the degrees of freedom of the right well. $S$ measures the bi-partite entanglement, that is the amount of genuine quantum correlations between the left well and the right one, seen as two partitions of the whole quantum system. The entanglement entropy $S$ is given by
\beqa
\label{ee}
S=-Tr \hat{\rho_{L}} \log_{2} \hat{\rho_{L}}=
-\sum_{i=0}^{N}|c_{i}^{(0)}|^2\log_{2}|c_{i}^{(0)}|^2
\;.\eeqa
We observe that the cat state is not the maximally entangled state - from the left-right bi-partition perspective - for our system. In fact, from Eq. (\ref{ee}) we see that, for a NOON state, $S=1$, while the maximum value attained by the entanglement entropy for $N=100$ is $S_{max}\simeq 5.4$, see panel d) Fig. \ref{fig1}.
From analytical considerations based on the asymptotic shape of the reduced density matrix
\cite{dellanna,lucamichele} we find that the peak of the entropy weakly depends on the interacion parameters and, in the large $N$ limit, is simply given by
\beq
S_{max}\simeq \frac{1}{2}\log_2(2\pi e N)\;,
\eeq
while the location of the peak is around $V=U+2/N$, in perfect agreement with the numerical results. The value of $S$ at $V=0$, in the large $N$ limit, is recovered \cite{dellanna}
\beq
S(V=0)\simeq \frac{1}{2}\log_2\left(\pi e \sqrt{\frac{N^2}{4+2 UN}}\right)\,,
\eeq
as well as its asymptotic value for
$V\rightarrow \infty$ which is $S\rightarrow 1$, sign of the NOON state.

$F$, $\alpha$, and $S$, as functions of $V$ are shown in Fig. 1. We see that in correspondence to an increasing of the nearest-neighbor interaction the ground-state of the system experiences a "catness" enhancement accompanied by a softening of the coherence visibility and of the entanglement entropy which gets its maximum value when $V=V_{cr}$. This trend can be retrieved also for larger values of $U$. For each of three indicators - $F$, $\alpha$, $S$ - in Fig. 1, we show the comparison for two different values of $U$. We have found - as expected from Eq. (\ref{linearrelation}) and the following discussion - that an increasing of the repulsive interatomic interaction produces a shift of $V_{cr}$ towards larger values. Notice that also when the cross-well interaction is absent, $V=0$, the entanglement entropy $S$ attains its maximum value when the ground-state of the underlying Hamiltonian is a cat-like state \cite{cats,dellanna,pierfrancesco}.

{
By means of long-range potential one could circumvent the collapse - and thus the instability - of the bosonic cloud which should take place with sufficiently high densities for attractive interactions. Indeed, the long-range potential between the confined bosons introduces an additional degree of freedom with respect to the short-range one - that is the nearest-neighbor density-density interaction - which when suitably tuned makes possible to get the NOON state even for repulsive interactions. We stress that such a phenomenon is absent in the standard two-site Bose-Hubbard model, i.e. in the absence of inter-well interactions \cite{huang,bruno,cats,dellanna,pierfrancesco,bruno2}}.

\section{Conclusions}

We have considered an atomic Josephson junction made of a finite number of interacting dipolar bosons. By employing the extended two-site Bose-Hubbard Hamiltonian, we have carried out the zero-temperature analysis by finding the ground-state of the system and characterizing it with the help of three indicators: the Fisher information, the coherence visibility, and the entanglement entropy. We have studied these quantities by fixing the on-site interaction and varying the nearest-neighbor one. We have found that the presence of a Schr\"odinger-cat like state in the double-well corresponds to sufficiently large values of the Fisher information. We have pointed out that the cat-like state emerges when, within the underlying classical model for the junction, the population imbalance parameter oscillates around a non-zero value. This kind of oscillations sets in for the nearest-neighbor interaction signing the maximum of the entanglement entropy. In this situation the coherence visibility is quite small but it increases as the inter-well interaction strength becomes sufficiently small.

As for what concerns the future perspectives, the bosonic dipolar interaction might be very useful in dynamical generating cat-like states with one \cite{gentaro} and two bosonic components \cite{adeleroberta} since it is able to make wider the separation with the first excited state, making in this way the cat state more robust against decoherence.
\\

The present work has been supported by Progetto Giovani (University of Padova): "Many Body Quantum Physics and Quantum Control with Ultracold Atomic Gases" and by Progetto di Ateneo (University of Padova): "Quantum Information with Ultracold Atoms in Optical Lattices".

\end{document}